\documentclass[12pt,preprint]{aastex}

\slugcomment{To be submitted to ApJ}

\shorttitle{Strong Lensing in the RCS}
\shortauthors{Gladders et al.}

\begin{document}

\title{The Incidence of Strong-Lensing Clusters in the \\
Red-Sequence Cluster Survey\altaffilmark{1}}

\author{Michael D. Gladders\altaffilmark{2,3}} \affil{The Carnegie
Observatories, Pasadena, CA 91101, USA}

\and

\author{Henk Hoekstra\altaffilmark{2}} \affil{The Canadian Institute
for Theoretical Astrophysics and Department of Astronomy \&
Astrophysics, U. of Toronto, Toronto, ON M5S 3H8, Canada}

\and

\author{H.K.C. Yee\altaffilmark{2,3}} \affil{Department of Astronomy
\& Astrophysics, U. of Toronto, Toronto, ON M5S 3H8, Canada}

\and

\author{Patrick B. Hall\altaffilmark{2,3}} \affil{Princeton University
Observatory, Princeton, NJ 08544 and Departamento de Astronom\'{\i}a y
Astrof\'{\i}sica, Pontificia Universidad Cat\'{o}lica de Chile,
Casilla 306, Santiago 22, Chile}

\and

\author{L. Felipe Barrientos\altaffilmark{3}} \affil{Departamento de
Astronom\'{\i}a y Astrof\'{\i}sica, Universidad Cat\'{o}lica de Chile,
Casilla 306, Santiago 22, Chile}

\altaffiltext{1}{Based on observations collected at the Baade 6.5m
Telescope, the CFHT 3.6m telescope, and the CTIO 4m telescope.}

\altaffiltext{2}{Visiting Astronomer, Canada-France-Hawaii Telescope,
which is operated by the National Research Council of Canada, le
Centre Nationale de la Recherche Scientifique, and the University of
Hawaii.}

\altaffiltext{3}{Visiting Astronomer, Cerro Tololo Inter-American Observatory.
CTIO is operated by AURA, Inc.\ under contract to the National Science
Foundation.}


\begin{abstract}
The incidence of giant arcs due to strong-lensing clusters of galaxies
is known to be discrepant with current theoretical expectations. This
result derives from a comparison of several cluster samples to
predictions in the framework of the currently favored $\Lambda$CDM
cosmology, and one possible explanation for the discrepancy is that
this cosmological model is not correct.  In this paper we discuss the
incidence of giant arcs in the Red-Sequence Cluster Survey (RCS),
which again shows significant disagreement with theoretical
predictions. We briefly describe a total of eight strong lens systems,
seven of which are discussed here for the first time.  Based on the
details of these systems, in particular on the ratio of single to
multiple arc systems, we argue that it may be possible to explain this
discrepancy in the currently favored cosmology, by modifying the
details of the lenses themselves. Specifically, the high incidence of
multiple arc systems and their overall high redshift suggests that a
sub-population of the global cluster population is responsible for
much of the observed lensing. The lack of lensing clusters at
$z<0.64$ in the RCS indicates that a property associated
with clusters at early times results in the boosted lensing
cross sections; likely a combination of ellipticity and elongation
along the line of sight, substructure, and changes in the cluster mass
profiles is responsible. Cluster mass, which should evolve to globally
higher values toward lower redshifts, is clearly not the most
significant consideration for the formation of giant arcs.

\end{abstract}




\section{Introduction}

The incidence of giant arcs due to the strong lensing effects of
galaxy clusters is in principle calculable, given a known cluster
population, a cosmology, and a source population \cite[][hereafter
B98]{bar98}. The cosmology has an effect on the expected result, both
because of its effect on angular diameter distances, and more
importantly by its effect on the lens population. The evolution of the
space density of massive clusters with redshift is strongly affected
by the parameters $\sigma_8$ and $\Omega_M$
\citep[e.g.,][]{ouk92,car97,bah98}, and to a lesser extent $\Omega_\Lambda$
\citep[e.g.,][]{hai01}. This has a strong effect on the expected
lensing incidence. Moreover, the cosmology affects the internal
details of the cluster lens population; significant sub-structure and
ellipticity are known to boost lensing cross sections
\citep[e.g.,][]{ogu02,bar95} and the occurrence of such features is
affected by cosmology \citep[e.g.,][]{ric92}. Unfortunately, the
internal properties of galaxy clusters are likely also affected by
non-cosmological physics, including the detailed nature of dark matter
\citep[e.g.,][]{men01}, and the effects of baryons on the cluster mass
profile in the cluster core \citep[e.g.,][]{wil99}. The source
population also affects the incidence of arcs.  These various
complications make the prediction of arc statistics a challenging
problem, and likely limits the cosmological impact of such studies.

However, B98 showed that the influence of cosmology on arc statistics
is dramatic, with variations in arc counts of several orders of
magnitude between extreme models. Given this sensitivity to cosmology,
one might hope that this effect would dominate and allow one to use
arc statistics as a cosmological test. Notably, both the results of
B98 using the EMSS (Einstein Medium Sensitivity Survey) cluster sample
and \cite{zar02} using primarily the LCDCS (Las Campanas Distant
Cluster Survey) cluster sample show that the observations clearly
favor an open, low-$\Omega_M$ universe. Both samples overproduce arcs
in comparison to the currently favored flat $\Lambda$CDM model by a
factor of $\sim$10.

In this paper we describe another cluster arc sample which has far too
many arcs compared to the standard expectation for a $\Lambda$CDM
cosmology. The cluster sample is drawn from the Red-Sequence Cluster
Survey \citep[RCS;][]{gla03a} and includes five clusters with arcs in a
primary sample drawn directly from the survey images and a further
three from a secondary followup sample. We briefly describe the survey and
each arc system in \S2. In \S3 we analyse the occurrence rates and
redshifts of single- and multiple-arc clusters and from that conclude
that internal lens structure, rather than cosmology, is the likely
cause of the discrepancy.

\section{The RCS Sample}

The RCS is a $\sim$90 square degree $R_C$- and $z'$-band imaging
survey designed primarily to search for galaxy clusters to redshifts
as high as $z=1.4$ \citep[see][for further details]{gla03a}. The
images, acquired at the CFHT and CTIO 4m telescopes using mosaic
cameras, are relatively shallow, with $5\sigma$ point source limits
typically 24.8 mag in $R_C$ (Vega normalized) and 23.6 mag in $z'$
(SDSS normalized). Despite this, the original survey contains 5
clusters with features interpreted as giant arcs due to strong
lensing. We designate this the primary sample. An ongoing follow-up
project, consisting in part of deeper $I$-band imaging using the Baade
6.5m telescope of one hundred of the most significant $z>0.95$ RCS
cluster candidates, has also turned up a secondary sample of 3
clusters with giant arcs, selected from a sample to date of 46
clusters. 

In both samples, the arcs were detected by visual examination of known
cluster fields, with the cluster found in each case as an over-density
of red galaxies using a variant of the algorithm defined in
\cite{gla00}. Several of the brightest and most obvious arcs in the
primary sample were found independently by a direct examination of
images during the initial observing. The effective selection area for
these brightest arcs in the primary sample is the full area of the
RC. For some fainter arcs variations in seeing and sky brightness make
the effective areas smaller; the images of clusters with arcs all have
seeing better than one arcsecond, and approximately 80\% of the RCS
has similar image quality. Overall, the survey data for fields with
arc clusters are typical, both in seeing and sky brightness. All of
the RCS survey data has also been visually examined by various
individuals to check the object finding in the primary processing, and
a large fraction of the good-seeing data were re-examined by
H. Hoekstra while checking all the weak-lensing results
\citep{hoe02}. It is thus highly unlikely that comparable undiscovered
arcs exist in the RCS survey data. 

The secondary arc sample is drawn from data with much better seeing
than the RCS data - a median seeing of 0\arcsec.45 for the run, with
no image worse than 0\arcsec.7 - and all these data have been visually
examined for arcs. No other arcs comparable to the three in the
secondary sample exist in these data.

Below we briefly describe each arc system, and argue the
strong-lensing interpretation for each object. A summary of the
cluster and arc properties, including length-to-width ratios for the
arcs, is given in Table 1. Table 1 includes a measure of the relative
rank of the detection significance for each cluster.  This is computed
considering all RCS cluster candidates over the redshift spanned by
all the lens candidates for the primary sample ($0.64<z<0.87$), and
within only the 46 clusters imaged so far when considering the
secondary sample. Table 1 also provides an estimate of the surface
brightness limits at 5$\sigma$, per square arcsecond, for the
detection image for each arc.

\subsection{Primary Sample}
\subsubsection{RCS0224.5-0002}
Even considering only the ground-based data \citep{gla02a}, this
spectroscopically confirmed $z=0.773$ cluster is the single most
obvious case of strong lensing in the RCS, with the most distant arc
spectroscopically confirmed at $z=4.8786$. Recent HST
imaging \citep{gla03} shows that the features visible in the
ground-based data probably correspond to images of three different
background sources, with two of the sources at a similar
redshift. Sources at a similar redshift are clustered significantly on
the distance scales relevant here, and so conservatively this
represents a two-arc system to the depth imaged in the RCS. In the
deeper HST observations, another fainter giant arc system - either a
third or fourth arc system depending on interpretation - is also
visible \citep{gla03}.

\subsubsection{RCS0348.8-1017}
This single arc, shown in Figure 1a, is apparently produced by a poor
cluster at a photometric redshift of 0.80$\pm$0.05. The lens is at a
lower significance than the lower limit for the primary RCS
catalog. The arc is blue compared to other nearby objects; this fact,
as well as the morphology and the object's location with respect to an
apparent poor cluster, argues for the lensing interpretation in this
case.

\subsubsection{RCS1324.5+2845}
This extremely rich cluster, at a photometric redshift of 0.85$\pm$0.05,
produces one apparent giant arc, and several other features which are
suggestive of strong lensing.  As can be seen in Figure 1b, follow-up
of this system is complicated by the presence of two nearby bright
stars. The interpretation of the extended feature as an arc is
primarily based on its extremely elongated morphology, and tangential
alignment with respect to the cluster center.

\subsubsection{RCS1419.2+5326}
RCS1419.2+5326, shown in Figure 1c, is a spectroscopically confirmed
$z=0.64$ cluster \citep{ell03} which produces two obvious giant arcs.  These objects
are both blue. The color, overall morphology, as well as the
tangential arrangement with respect to the cluster core suggests a
strong lensing interpretation. Moreover, note that the fainter of the
two putative arcs is about twice as distant from the cluster center as
the brighter arc. Giant arcs occur near critical curves, and the
significant differences in radial position of these arcs makes it
unlikely that they are at similar redshifts. The more distant arc also
has a lower surface brightness, and it is also marginally redder.
Based on these differences, we suggest that the two arcs correspond to
images of two different background sources, with the fainter
corresponding to a source at significantly higher redshift than the
brighter. 

\subsubsection{RCS1620.2+2929}
This spectroscopically confirmed $z=0.87$ compact and rich cluster \citep{ell03} has
one feature which is suggestive of an arc, and several other smaller
sources with strong tangential shear apparently aligned about the
cluster core. This arc candidate, shown in Figure 1d, has an unusual
color compared to all other galaxies in the immediate field. The color
of the galaxy as well as its morphology are consistent with a
strong-lensing interpretation, but this should be considered the least
secure arc candidate in the primary sample.

\subsection{Secondary Sample}

\subsubsection{RCS2122.9-6150}
This cluster is shown in Figure 2a, and is at a photometric redshift of
1.1$\pm$0.10. The giant arc candidate is obvious, consisting of an extended
feature near and somewhat tangential to the apparent cluster cD. This
object is not visible in the original RCS survey data and hence no
useful color information is currently available. 
\subsubsection{RCS2156.7-0448}
RCS2156.7-0448, shown in Figure 2b, is a rich cluster at a photometric
redshift of 1.2$\pm$0.10. It has one candidate arc, located near the
apparent cluster cD galaxy and tangential to it. This object also has
a relatively uniform surface brightness along its entire length,
arguing against it being a projection of several disk galaxies. The
putative arc is barely visible in the original RCS survey imaging, and
appears bluer than the cluster galaxies in that data. This should be
considered the least-likely arc candidate in the secondary sample.
\subsubsection{RCS2319.9+0038}
This rich cluster is at a photometric redshift of
1.0$\pm$0.1. $I$-band Baade 6.5m imaging, shown in Figure 2c, suggests
the presence of two arcs. Bluer imaging, shown in Figure 2d, reveals
that this is in fact a three arc system, and clearly confirms that
RCS2319.9+0038 is another spectacular example of strong lensing by a
high redshift cluster, comparable to RCS0224.5-0002. These data also
show that one of the arcs is a $B$-band dropout. For the purposes of
constructing a statistical sample we consider only the $I$-band data
and treat this as a two arc system.

\begin{deluxetable}{llccccl}
\tabletypesize{\scriptsize}
\tablecaption{Basic Parameters of RCS Lensing Clusters and Associated Arcs. \label{tbl-1}}
\tablewidth{0pt}
\tablehead{
\colhead{Cluster Name} & \colhead{Redshift} & \colhead{Surface Brightness}  & \colhead{Rank\#} & \colhead{Arc Redshift(s)} & \colhead{Arc $l/w$}&Notes}
\startdata
RCS0224.5-0002   & 0.773         &23.82   & 1      & $\sim1.7$ & $>24$    &3rd(or 4th) fainter arc\\
                 &               &        &        & 4.8786    & $>28$    &~~~~in HST imaging\\ 
RCS0348.8-1017   & 0.80$\pm$0.05 &23.63   &$>1752$\tablenotemark{a} & ------    & $>8$     & \\
RCS1324.5+2845   & 0.85$\pm$0.05 &23.95   & 98     & ------    &$\sim13$  &\\
RCS1419.2+5326   & 0.64          &23.97   & 16     & ------    &$\sim11$  &\\
                 &               &        &        & ------    &$>8$      &\\
RCS1620.2+2929   & 0.87          &23.93   & 20     & ------    & $>5$     &tentative ID\\\hline
RCS2122.9-6150   & 1.20$\pm$0.10 &24.10   & 31     & ------    &$\sim15$  &\\
RCS2156.7-0448   & 1.10$\pm$0.10 &24.11   & 43     & ------    &$\sim6$   &tentative ID\\
RCS2319.9+0038   & 1.00$\pm$0.05 &23.90   & 1      & ------    & $\sim12$ &3rd arc visible in\\
                 &               &        &        & $3-4$     & $\sim7$  &~~~~bluer imaging\\
\enddata


\tablecomments{Lens redshifts with error bars are photometric
estimates, and are otherwise spectroscopic redshifts. Arc
length-to-width ratios, $l/w$, are reported as lower limits if the arc
appears unresolved. Arc redshifts are either spectroscopic,
photometric, or based on lens modeling. The quoted surface
brightnesses are 5$\sigma$ magnitude limits per square arcsecond for the detection
image for each arc: $R_C$-band for the RCS primary sample, and
$I_C$-band for the secondary sample, Vega calibrated using
\cite{lan92} standards in both cases. Uncertainties are less than 0.1
magnitudes.} \tablenotetext{a}{The apparent poor cluster producing
this arc does not appear in the primary RCS cluster sample, since its
significance is below the threshold.}

\end{deluxetable}

\section{Discussion}
The standard prediction from B98 for the number of
giant arcs due to clusters in a $\Lambda$CDM universe is approximately
one arc per 150 square degrees, integrated over $0<z<1$, for sources
at $z=1$. Though the RCS cluster sample probes a somewhat different
redshift range (in particular the secondary arc sample described
above) the resulting differences should be factors of order
unity. Even considering only the primary RCS arc sample (5 clusters, 7
arcs) and ignoring details of specific arc length-width sub-samples, we
find disagreement with the Bartlemann $\Lambda$CDM predictions by a
factor of 10-20. Similar disagreements exist for the EMSS
sample (B98)  and the LCDCS \citep{zar02}.

There are three basic ways to increase the number of arcs predicted:
increase the surface density of sources, increase the number of
lenses, or increase the cross section of the lenses. Cosmology affects
the latter two, but not the first. The number of sources, and their
redshift distribution, is a well established observable, and is now in
principle known to extremely faint limits due to the Hubble Deep
Fields \citep[e.g.][and references therein]{cas00}. The expected
number of lenses is a strong function of the cosmology, and observably
depends on the mass limit in a given sample. Establishing precisely
the same limit in theoretical calculations is non-trivial and may
represent a significant source of error in comparing observations to
predictions.  The cross section of individual lenses is also a
function of mass, with more massive clusters having larger cross
sections. Moreover, lensing cross sections can be strongly affected by such
things as lens ellipticity \citep{ogu02}, projected secondary
structures \citep{wyi01}, internal cluster substructure \citep{bar95},
the presence of a central cD galaxy \citep{wil99}, and the cluster
mass profile \citep{tak01}, at least some of which are also affected
by cosmology \citep[e.g.,][]{ric92}. The cross section of a given lens
for a source at a given redshift is also a function of cosmology since
this affects the size of the lens caustics in the source plane.

A further expression of the discrepancy between theory and
observations is suggested by the relative proportion of single to
multiple arc clusters. In either the RCS primary or secondary sample,
in which arcs are found on the basis of examining images of uniform
depth of a large number of clusters, the probability of two arcs
occurring around any one cluster is approximately $P^2$, if the
probability of forming a single arc is merely $P$. This ignores the
effects of source redshift on the lensing cross section, but is of
sufficient precision for the following discussion.  If one makes the
further simple assumption that all the lensing clusters are drawn from
the same parent population, each member of which has a similar lensing
cross-section, the implication from the primary sample is that the
probability, $P$, of forming an arc for any one cluster is
2/5. Similarly $P=1/3$ from the secondary sample. Both samples also
contain one single-arc system with a tentative identification;
exclusion of these systems implies even higher lensing probabilities.
However, based on the detection rankings in Table 1, it appears that
many of the individual clusters with arcs do not stand out in the
context of the whole cluster sample, with many other clusters without
arcs showing similar or greater cluster detection significance. 

To produce the large proportion of multiple arc clusters seen in the
RCS, we require at least some clusters with large lensing
probabilities, and this conclusion is independent of the number of
clusters considered. Given the large number of clusters in the RCS
which do not show arcs, it seems likely that the distribution of
lensing probabilities is strongly skewed, with a tail of high
probability lenses. Mass is an obvious underlying property which
might cause this, since variations in the cluster mass as given by the
cluster mass function produces a small sub-population (the most
massive clusters) which have an enhanced lensing cross section.

To investigate the effect of the cluster mass function on the expected
proportion of single to multiple arc clusters, we use the cluster
catalogs from the Hubble Volume Virgo simulations\footnote{The
simulations used in this paper were carried out by the Virgo
Supercomputing Consortium using computers based at the Computing
Centre of the Max-Planck Society in Garching and at the Edinburgh
Parallel Computing Centre. The data are publicly available at \texttt
{http://www.mpa-garching.mpg.de/NumCos}} \citep{evr02} to construct
mock samples of clusters over the redshift and mass range explored by
the RCS. In simple symmetric lens models the probability of forming
giant arcs for any one cluster scales linearly with mass because the
length of the caustic is proportional to mass; to form a giant arc one
must cross the caustic, and so length is the relevant
quantity. Detailed measurements of known strong lensing clusters
suggest that the relationship between mass and lensing power is
shallower than this \citep{wil99}, though with significant
scatter. Simulations conversely suggest that the scaling between mass
and lensing cross section is steeper \citep[e.g.,][]{men03}. We use
the simplest model, with the lensing cross section proportional to
mass, as a middle ground between these extremes. This is sufficient
for the simple illustrative models shown here.

The lensing probability for each cluster in the mass function is set
relative to a fiducial value, $P_{14.7}$. $P_{14.7}$, which is the
lensing probability for a 5$\times10^{14}$h$^{-1}$M$\odot$ cluster, is
arbitrarily adjusted in order to reproduce the number of single arc
systems observed in the primary sample. This particular mass is chosen
since it is approximately the lower mass limit used by B98 when
comparing to arc numbers in the EMSS cluster sample \citep{lef94}.

The value of $P_{14.7}$ is a function of the lower mass limit used,
since a lower mass limit yields a larger number of clusters in the
mock catalog, and hence requires a lower values of $P_{14.7}$ in order
to reproduce a fixed number of single arcs. Figure 3 shows the value
of $P_{14.7}$ required to reproduce the number of single arc clusters
observed, versus mass limit for limits ranging from
0.5$\times10^{14}$h$^{-1}$M$_\odot$ to
5.5$\times10^{14}$h$^{-1}$M$_\odot$. The resulting percentage of
multiple to single arc clusters is also shown. Limits greater than
5.5$\times10^{14}$h$^{-1}$M$_\odot$ produce a sample with an
insufficient number of clusters and are hence not considered. At any
reasonable mass cut, this model $overproduces$ arcs compared to the
results of B98, and always $underproduces$ multiple-arc clusters as
observed by the RCS. Overall, simply scaling the cluster lensing cross
section by the cluster mass produces a poor fit both to the RCS
observations and previous extensive modeling efforts.

A better match to both the results of B98 and the RCS than that shown
in Figure 3 is achieved if the distribution of probabilities is even
more skewed than one would infer from the cluster mass function.  As a
simple test, we consider an ad hoc model in which a small fraction of
clusters have a dramatically increased lensing cross section. Such a
modification has the advantage of keeping the lensing probabilities
for the bulk of the cluster fairly low. B98 modeled only a small
sample of clusters (only nine in any one cosmology) and might well not
have included any of these suggested extreme systems. Note also that
changes in source populations and cosmology may do little to produce
changes in the $distribution$ of lensing probabilities. Effects
producing a global increase in the lensing probabilities will enhance
the ratio of multiple to single arcs found, but at the expense of
worsening the already poor agreement with the B98 $\Lambda$CDM result.

Figure 4, similar to Figure 3, shows the result of a particular toy
model in which a random ten percent of all clusters have lensing cross
sections increased by a factor of ten. Figure 4 considers the lensing
probabilities both for the entire cluster population and only the
``typical'' clusters; the former produces overall lensing
probabilities consistent with the EMSS as seen by \cite{lup99}, while
the latter produces probabilities much closer to those modeled by
B98. The percentage of multiple arc clusters produced is now
significantly higher, with values not inconsistent with those seen
from the primary and secondary RCS samples.  Notably, further
observations of the two most striking double-arc clusters in this
sample also suggest that these clusters are remarkably good lenses,
and hence that positing a sub-population of ``super lenses'' is not
unreasonable.  As discussed in \S2.2.1, bluer and deeper observations
of both clusters reveal further giant arc features in each system. In
particular, RCS0224-0003 shows what can be interpreted as three giant
arc systems in the initial ground-based images, and clearly shows a
fourth giant arc system in relatively shallow HST imaging
\citep{gla03}.

The redshift distribution of the RCS primary arc sample provides
conclusive evidence that, as suggested by the above toy models, mass
is not the dominant factor which determines lensing cross
sections. For the more massive clusters which make up the bulk of the
parent population of lensing clusters, the RCS is complete from about
$0<z<1.1$ \citep{gla03a}. However, all the RCS clusters in the primary
sample are at $0.64<z<0.87$, despite the fact that clusters at
moderate redshifts($z\sim$0.3-0.4) are better lenses because their
caustics are bigger for distant sources, and that massive clusters are
more abundant at lower redshift. Qualitatively, based on the Hubble
Volume simulations used above, and taking
$M\geq$5.0$\times10^{14}$h$^{-1}$M$_\odot$ as a cut for massive
clusters, the median redshift of massive clusters is only 0.49 in a
$\Omega_M=0.3$, $\Omega_\Lambda=0.7$ universe, and less than 30\% of
all such clusters are at a redshift higher than the lowest redshift
arc cluster seen in the RCS primary sample.  The probability that the
RCS primary arc cluster sample is drawn from this population of
simulated massive clusters is less than 0.001. Something must act to
reduce the cross section of analogs to the RCS primary sample clusters
at later times.

Observations of clusters at redshifts similar to the RCS arc
clusters indicate that massive high-redshift clusters are often
elongated or occur with associated superstructure
\citep[e.g.,][]{gio99} and likely have significant substructure; seen
in appropriate projection such systems will have enhanced lensing
cross sections.  If the formation process of clusters tends to produce
more such systems at higher redshifts, this might explain both the
high proportion of multiple arc clusters seen, and their tendency to
be at unexpectedly high redshifts. 

Another possible explanation of the redshift distribution of the
RCS lenses is that the cluster potentials are more concentrated at
high redshift . Such an effect is well known from n-body simulations,
in which the concentration of a given halo is observed to be
correlated with the value of $\Omega_M$ at the redshift at which it
collapsed \citep{nav97}. The RCS sample may be the first sample with
sufficient redshift grasp to observe this effect. Moreover, the
multiple-arc clusters, particularly RCS0224-0003, have regular
concentric arcs; this implies that a single potential is responsible
for the lensing, and that it is unlikely to be highly
substructured. This contrasts with the multiple arc systems seen in
deep HST imaging around extremely massive lower redshift clusters
\citep[e.g.][]{kne96} in which substructure clearly has a significant
effect.

Finally, we note that in general, better agreement is found with the
results of B98 by comparing the RCS samples to an open CDM model,
since this model produces approximately the correct number of
arcs. The results of B98 show that a significant portion of this
increase in arcs numbers is due to larger lensing cross sections, in
addition to a global increase in the number of clusters in such a
cosmology. It is unclear whether this may result in the skewed lensing
probability distribution required by the RCS data, and whether an open
CDM model would correctly reproduce the multiple to single arc ratio
observed. It would be useful to redo the calculations of B98 with this
in mind, in particular paying attention to multiple arc
systems. Though numerous other observations appear to indicate that
open CDM models are unlikely, and in particular the CMB results
indicate that the universe is near to flat \citep[e.g.][]{deb02}, the
continuing suggestion that an open CDM universe is preferred by arc
statistics makes this topic worth revisiting.

\section{Conclusions}
We have presented a total of seven new lensing clusters from the
Red-Sequence Cluster Survey. In conjunction with one system already in
the literature \citep{gla02a}, this sample has been analysed in the
context of theoretical predictions for lensing statistics in a
$\Lambda$CDM universe. The long-standing disagreement between theory
and observations, in which the actual arc numbers are severely
under-predicted for this cosmology, is confirmed by these new
data. The most striking property of the RCS sample is the large number
(3 of 8) of multiple arc lensing clusters seen. An open CDM model is
still preferred by the RCS data, since this produces more arcs, in
part due to clusters with individually larger cross sections and thus
a greater tendency to produce multiple arc systems.

The high frequency of multiple arc systems result implies
that there exists a sub-population of clusters which are
extraordinarily good lenses. The overall high redshift of the RCS lens
sample suggest that the source of these ``super-lenses'' is likely
related to the process of cluster formation, and possibly due to some
combination of substructure, lens ellipticity, and projection of
associated structure along the line of sight. An alternate
interpretation is that these lenses are particularly dense, as is
expected on theoretical grounds for clusters which form
early. Regardless of cause, these results suggest that some physical
effect must serve to reduce the lensing cross sections of clusters at
later times, the opposite of what is expected if mass is the primary
parameter controlling lensing cross sections. 

Notably, these effects may have not been well captured in previous
modeling efforts due to the small number of clusters simulated, and an
incomplete treatment of surrounding structures. We suggest that
future efforts to model lensing by clusters must include both a large
number of clusters, and the complete line of sight to each in order to
correctly model such effects.  Consideration of multiple arcs in such
models may help rehabilitate arc statistics as a cosmological tool, as
it provides an independent check on the modeling.  Finally, we note
that the presence of a skewed distribution of lensing probabilities
for clusters, implied by our data, may complicate the use of such
clusters as probes of the global properties of cluster dark matter
haloes. Clusters showing arcs may represent a significantly biased
sample. 

\acknowledgments We thank the CFHT, CTIO and Magellan TACs for
generous allocations of observing time. MDG is partially supported by
the Natural Sciences and Engineering Research Council (NSERC) of
Canada via an NSERC PDF, and thanks G. Oemler for useful
discussions. MDG is also partially supported by an HST GTO grant. PBH
acknowledges financial support from Fundaci\'{o}n Andes. LFB's
research is supported by Fondecyt under proyecto \#1000537. The
research of HY is supported by grants from NSERC and the University of
Toronto.




\onecolumn

\figurenum{1}
\epsscale{0.8}
\begin{figure}[htb]
\plotone{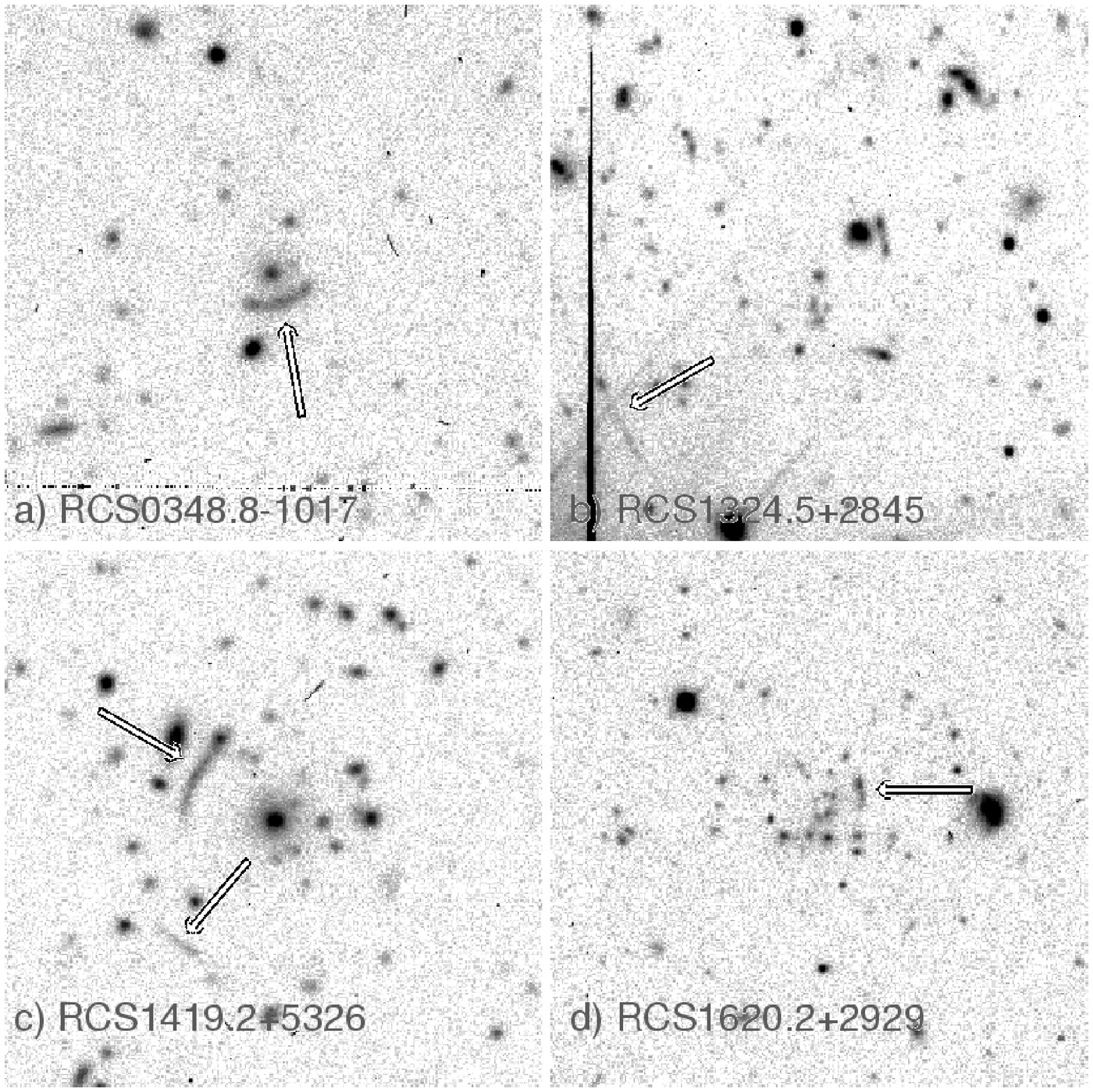}
\caption{{\footnotesize The four panels show greyscale $R_C$-band
images of the central 1'$\times$1' of the four new lensing clusters in
the primary sample discussed in \S2.1. The putative arc features are
indicated by an arrow in each case.
}}
\end{figure}

\clearpage
\onecolumn

\figurenum{2}
\epsscale{0.8}
\begin{figure}[htb]
\plotone{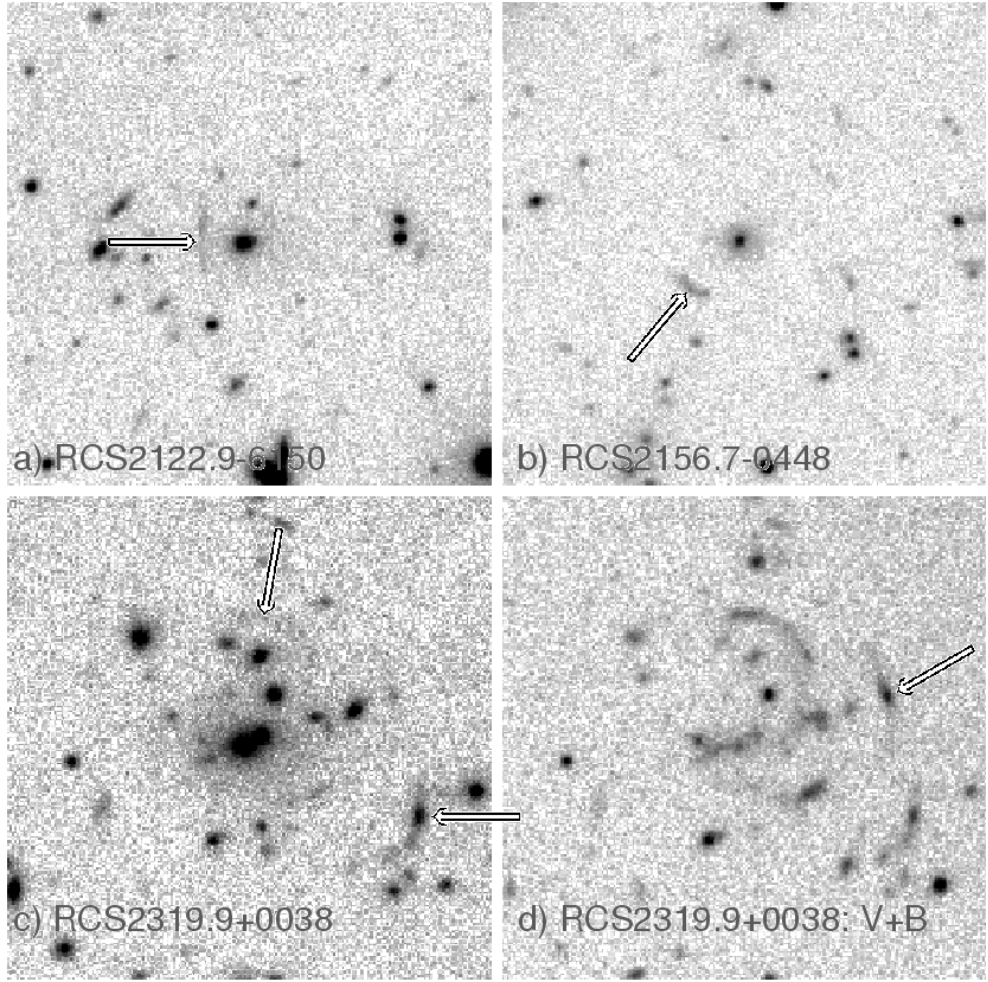}
\caption{{\footnotesize Panels a-c show greyscale $I$-band images of
the central 30''$\times$30'' of the three new lensing clusters in the
secondary sample discussed in \S2.2. The putative arc features are
indicated by an arrow in each case. Panel d shows RCS2319.9+0038 in
$V$ and $B$ light, from a summed image in which the arcs have similar
S/N in each filter. The $B$-band dropout indicated in the main text is
the arc to the bottom right. A third arc apparent in these bluer data
is also indicated.}}
\end{figure}

\clearpage
\figurenum{3}
\epsscale{0.8}
\begin{figure}[htb]
\plotone{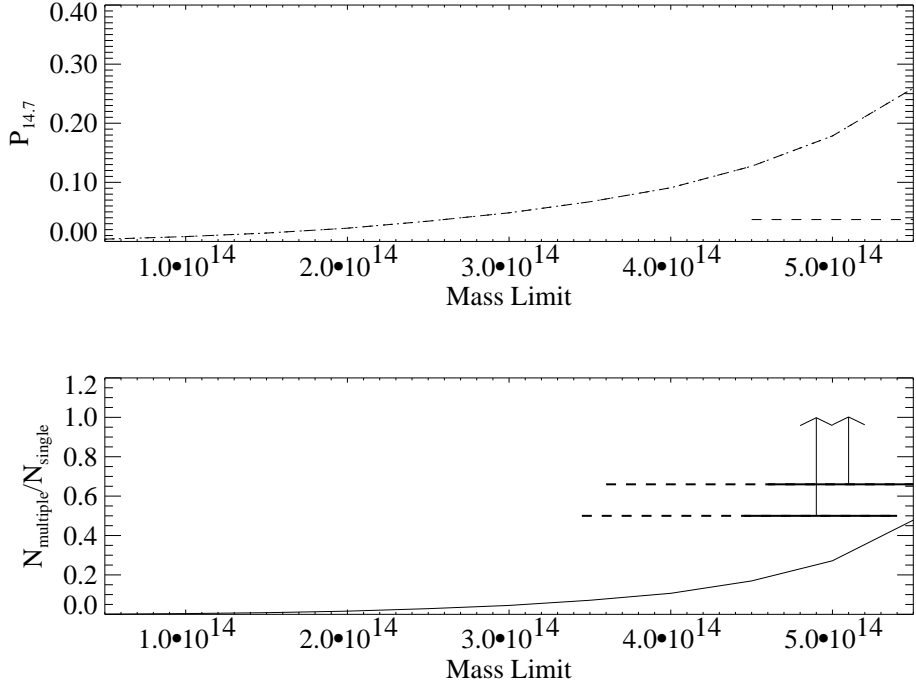}
\caption{{\footnotesize In the top panel, the dot-dashed line shows
the value of $P_{14.7}$ required to reproduce the number of single
arcs seen in the RCS primary sample, as function of mass limit, for a
model in which the cluster lensing probability scales linearly with
mass. The horizontal dashed line show the typical lensing probability
for a $\Lambda$CDM universe modeled by B98. As discussed extensively
in the main text, the RCS single arc data, and hence this model which
matches it, are inconsistent with the predictions of B98. The solid
line in the bottom panel shows the expected proportion of multiple to
single arc clusters for the same model. Horizontal solid lines show
the proportion of double-arc clusters seen in the RCS primary sample
(offset right) and the RCS secondary sample (offset left); it is
unclear what mass limit is appropriate for these clusters and this
uncertainty is suggested by the broken horizontal extension of these
lines to lower mass limits. Regardless, the model fails to match the
data at any mass limit. The effect of removing one of the single arc
clusters from each sample (possible because each sample contain one
tentative system with a smaller length-to-width ratio - see Table 1)
is also shown by the arrows.}}
\end{figure}

\figurenum{4}
\epsscale{0.8}
\begin{figure}[htb]
\plotone{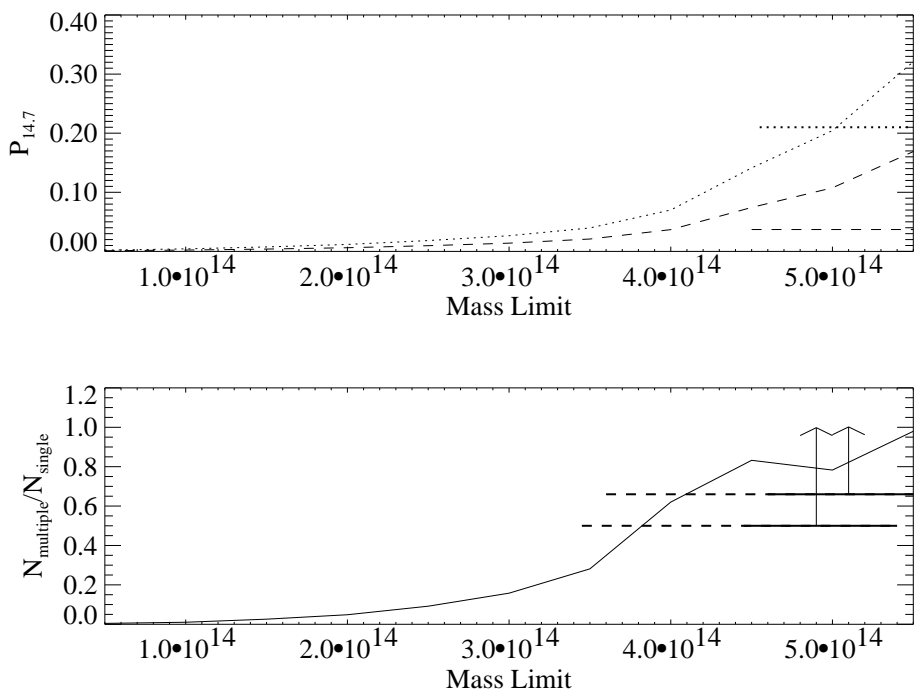}
\caption{{\footnotesize Similarly to Figure 3, except for a model in
which a random 10\% of all clusters have their lensing probabilities
boosted by a factor of 10. The value of $P_{14.7}$ is now shown for
the entire cluster sample (dotted line) and only the more typical
clusters (dashed line). The empirical probability of arcs for the EMSS
sample is shown by the horizontal dotted lines, and matches the model
result for the entire cluster sample quite well; the appropriate mass
limit for these data is also uncertain but likely at the massive end
of the range shown \citep{lup99} . Compared to Figure 3, this model
produces much better agreement with the observed ratio of multiple to
single arcs, while at the same time being more consistent with the
predictions of B98 for typical clusters.}}
\end{figure}


\begin{thebibliography}{}

\bibitem[Bahcall \& Fan(1998)]{bah98} Bahcall, N. A., \& Fan, X. 1998, \apj, 504, 1

\bibitem[Bartelmann et al.(1998)]{bar98} Bartelmann, M., Huss, A.,
Colberg, J.M., Jenkins, A., \& Pearce, F.R. 1998, A\&A, 330, 1 

\bibitem[Bartelmann, Steinmetz, \& Weiss(1995)]{bar95} Bartelmann, M.,
Steinmetz, M., Weiss, A. 1995, A\&A, 297, 1 

\bibitem[Carlberg et al.(1997)]{car97} Carlberg, R.G., Morris, S.,
Yee, H.K.C., \& Ellingson, E. 1997, \apj, 479, L19

\bibitem[Casertano et al.(2000)]{cas00} Casertano et al. 2000, \aj, 120,2747

\bibitem[de Bernardis et al.(2002)]{deb02} de Bernardis et al. 2002, \apj, 564, 559

\bibitem[Kneib et al.(1996)]{kne96} Kneib, J.-P., Ellis, R.S., Smail, I., Couch, W.J., \& 
Sharples, R.M. 1996, \apj, 471, 643

\bibitem[Ellingson et al.(2003)]{ell03} Ellingson, E. et al. 2003, in
preparation

\bibitem[Evrard et al.(2002)]{evr02} Evrard, A.E., MacFarland, T.,
Couchman, H.M.P., Colberg, J.M., Yoshida, N., White, S.D.M., Jenkins, A.,
Frenk, C.S., Pearce, F.R., Efstathiou, G., Peacock, J.A., \& Thomas, P.A.
2002, \apj, 573, 7

\bibitem[Haiman, Mohr, \& Holder(2001)]{hai01} Haiman, Z., Mohr, J.J., \& Holder, G.P. 2001, \apj, 553, 545

\bibitem[Hoekstra et al.(2002)]{hoe02} Hoekstra, H., Yee, H.K.C.,
Gladders, M.D., Barrientos, L.F., Hall, P.B., \& Infante, L. 2002,
\apj, 572, 55

\bibitem[Gioia et al.(1999)]{gio99} Gioia, I.M., Henry, J.P., Mullis, C.R., Ebeling, H., \& 
Wolter, A. 1999, \aj, 117, 2608

\bibitem[Gladders \& Yee(2003)]{gla03a} Gladders, M.D., \& Yee, H.K.C. 2003, in preparation

\bibitem[Gladders et al.(2003)]{gla03} Gladders, M.D., Hoekstra, H., Ellingson, E. \& Yee, H.K.C. 2003, in preparation

\bibitem[Gladders, Yee, \& Ellingson(2002)]{gla02a} Gladders, M.D., Yee, H.K.C., \&
Ellingson, E. 2002, \aj, 123, 1

\bibitem[Gladders \& Yee(2000)]{gla00} Gladders, M.D., \& Yee, H.K.C. 2000, \aj, 121, 2148

\bibitem[Landolt(1992)]{lan92} Landolt, A.U. 1992, \aj, 104, 340

\bibitem[Le F\`{e}vre et al.(1994)]{lef94}  Le F\`{e}vre, F., Angonin, M.C., Gioia, I.M., \& Luppino,G. A. 1994, \apjl, 422, 5

\bibitem[Luppino et al.(1999)]{lup99} Luppino, G.A., Gioia, I.M.,
Hammer, F., Le F\`{e}vre, O., Annis, J.A., 1999 A\&AS, 136, 117

\bibitem[Meneghetti et al.(2001)]{men01} Meneghetti, M., Yoshida, N.,
Bartelmann, M., Moscardini, L., Springel, V., Tormen, G., White,
S.D.M. 2001, \mnras, 325, 435

\bibitem[Meneghetti, Bartelmann \& Moscardini(2003)]{men03} Meneghetti, M.,
Bartelmann, M., \& Moscardini, L. 2003, \mnras, 340, 105

\bibitem[Navarro, Frenk, \& White(1997)]{nav97} Navarro, J.F.,
Frenk, C.S., \& White, S.D.M. 1997, \apj, 490,493

\bibitem[Oguri(2002)]{ogu02} Oguri, M. 2002, \apj, 573, 51

\bibitem[Oukbir \& Blanchard(1992)]{ouk92} Oukbir, J., \& Blanchard, A. 1992, A\&A, 262, 21O
 
\bibitem[Richstone, Loeb, \& Turner(1992)]{ric92} Richstone, D., Loeb, A., \& Turner, E.L. 1991, \apj, 393, 477

\bibitem[Takahashi \& Chiba(2001)]{tak01} Takahashi, R., \& Chiba, T. 2001, \apj, 563, 489

\bibitem[Williams, Navarro, \& Bartelmann(1999)]{wil99} Williams,
L.L.R., Navarro, J.F., \& Bartelmann, M. 1999, \apj, 527, 535

\bibitem[Wyithe, Turner, \& Spergel(2001)]{wyi01} Wyithe, J.S.B.,
Turner, E.L., \& Spergel, D.N. 2001, \apj, 555, 504 

\bibitem[Zaritsky \& Gonzalez(2002)]{zar02} Zaritsky, D., \& Gonzalez, A.H. 2002, \apj, in press, astro-ph/0210352

\end{thebibliography}
\end{document}